\documentclass[twocolumn,showpacs,preprintnumbers,amsmath,amssymb]{revtex4}

\usepackage{graphicx}
\usepackage{dcolumn}
\usepackage{bm}

\begin{document}

\preprint{APS/123-QED}

\title{Nonequilibrium evolution thermodynamics of vacancies}

\author{Leonid S. Metlov}

 \email{lsmet@fti.dn.ua}
\affiliation{Donetsk Institute of Physics and Engineering, Ukrainian
Academy of Sciences,
\\83114, R.Luxemburg str. 72, Donetsk, Ukraine
}%

\date{\today}

\begin{abstract}
An alternative approach - nonequilibrium evolution thermodynamics, is compared with classical Landau approach. A statistical justification of the approach is carried out with help of probability distribution function on an example of a solid with vacancies. Two kinds of kinetic equations are deduced in terms of the internal energy and the modified free energy. 
\end{abstract}

\maketitle


A great number of works devoted to development of the phase-field approaches for description of different nature phenomena has been published recently \cite{akv00,kkl01,esrc02,lpl02,rbkd04,gptwd05,akegny06,s06,rjm09}. They are continuing the Landau theory of phase transitions \cite{lh54,ll69}. During the last decade an alternative approach, close to the Landau technique of phase transitions, has been developing too \cite{mm97,m07a,m09,m10}. There is no any statistical justification of the approach and in the present article such justification is carried out with help of probability distribution function (PDF) on the example of a solid with vacancies. In the approach the conjugated pair of thermodynamic potentials, that is, the internal and (modified) free energy, is introduced. The main intrigue is that the equilibrium or stationary state does not coincide with extremes of such thermodynamic potentials \cite{m09,m10}.

\begin{figure}
\includegraphics [width=2.9 in]{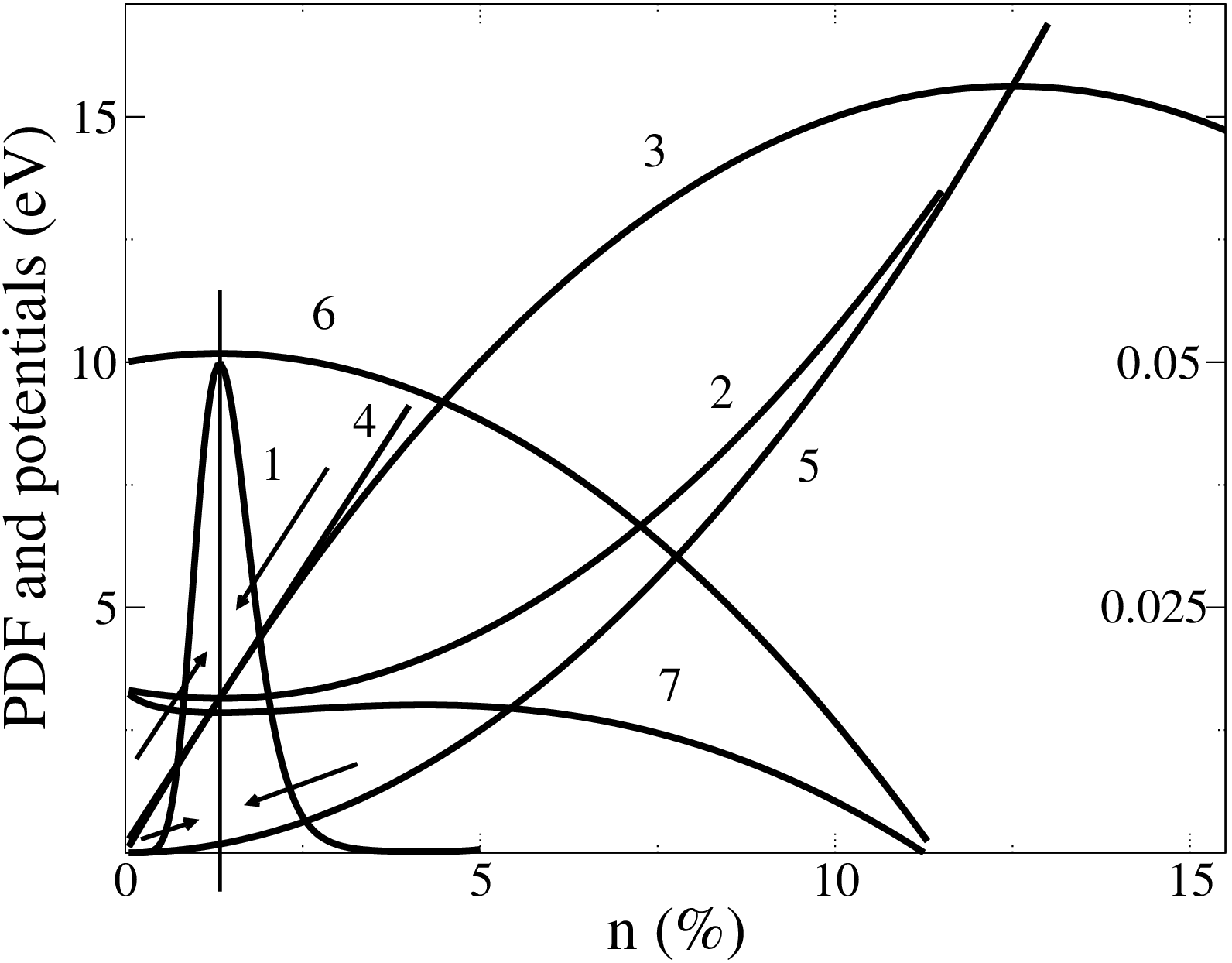}
\caption{\label{f1} Basic set of thermodynamic functions for quadratic approximation: 1 -- probability distribution function $f$; 2 -- classic configurational free energy $F_{c}$; 3 -- internal energy $U$; 4 -- a tangent to $U$ at the equilibrium point; 5 -- modified configurational free energy $\tilde{F_{c}}$; 6 -- effective internal energy $\bar{U}$; 7 -- effective configurational free energy $\bar{F_{c}}$. The right-hand scale is for PDF.}
\end{figure}

If a solid consisting of $N$ particles has $n$ point defects (e.g., vacancies, substituted atoms, etc.) then in this case the equilibrium (or stationary) state can be found from the maximum of PDF taken in the form \cite{f55, ll69, s89, g97} (curve 1 in Fig. \ref{f1})
  \begin{equation}\label{b1}
f(n)=CW\exp(-\dfrac{U(n)}{kT}),
  \end{equation}
where $C$ is a normalizing factor, $U(n)$ is the internal energy dependent on the number of structural defects, $k$ is the Boltzmann's constant and $W$ is the thermodynamic probability as a possible combination of the number of defects $n$ and the number of atoms $N$ of a crystal \cite{f55}
  \begin{equation}\label{b2}
W=\dfrac{(N+n)!}{N!n!},
  \end{equation}
the logarithm of which is a configurational entropy (curve 2 in Fig. 1)
  \begin{equation}\label{b3}
S_{c}=k\ln W.
  \end{equation}
Note, that the configurational entropy is a one-valued function of the number of defects. It is absolutely independent of the energy of defects (and of temperature too).

The pre-exponential factor of (\ref{b1}) describes the combinational, that is entropic, part of the distribution function, related to degeneration of macrostates. The exponential factor describes the restrictive part of the distribution function, connected with the overcoming of potential barriers between microstates. In a power presentation
  \begin{equation}\label{b4}
U=U_{0}+ \sum_{k=1}^{K} \frac{(-1)^{k-1}}{k}u_{k-1}n^k   ,
  \end{equation}
where $U$ is the internal energy of a solid with defects, $U_{0}$ is the internal energy of the defect-free solid, $u_{k-1}$ are some coefficients, $K$ is the highest power taken into account. In a linear approach $U=U_{0}+u_{0}n$ the same potential was introduced by Frenkel \cite{f55}. In a quadratic approximation the graph of $U$ is shown in Fig. \ref{f1} (curve 3).

Bringing the variables independent of the number of defects $n$ into the ``inessential'' constant $C$ we can write the expression (\ref{b1}) in the form
  \begin{equation}\label{b5}
f(n)=C\dfrac{(N+n)!}{n!}\exp(-\dfrac{U(n)}{kT}),
  \end{equation}
or as a product
  \begin{equation}\label{b6}
f(n)=C\prod_{i=1}^{N}(n+i)\exp(-\dfrac{U(n)}{kT}).
  \end{equation}

Through differentiation we obtain
  \begin{equation}\label{b7}
\frac{\partial f(n)}{\partial n}=(\sum_{k=1}^{N+n}\frac{1}{k}-\sum_{k=1}^{n}\frac{1}{k}-\frac{u}{kT})f(n),
  \end{equation}
where
  \begin{equation}\label{b7a}
u\equiv \frac{\partial U(n)}{\partial n}= \sum_{k=0}^{K-1} (-1)^{k}u_{k}n^k 
  \end{equation}
is the energy of a defect (tangent 4 to the internal energy in Fig. \ref{f1}).

The extreme value of probability distribution function is at n, it obeys the following transcendental equation
  \begin{equation}\label{b8}
\sum_{k=1}^{N+n}\frac{1}{k}-\sum_{k=1}^{n}\frac{1}{k}-\frac{u}{kT}=0.
  \end{equation}

\begin{figure}
\hspace{0.06 cm}
\includegraphics [width=3 in] {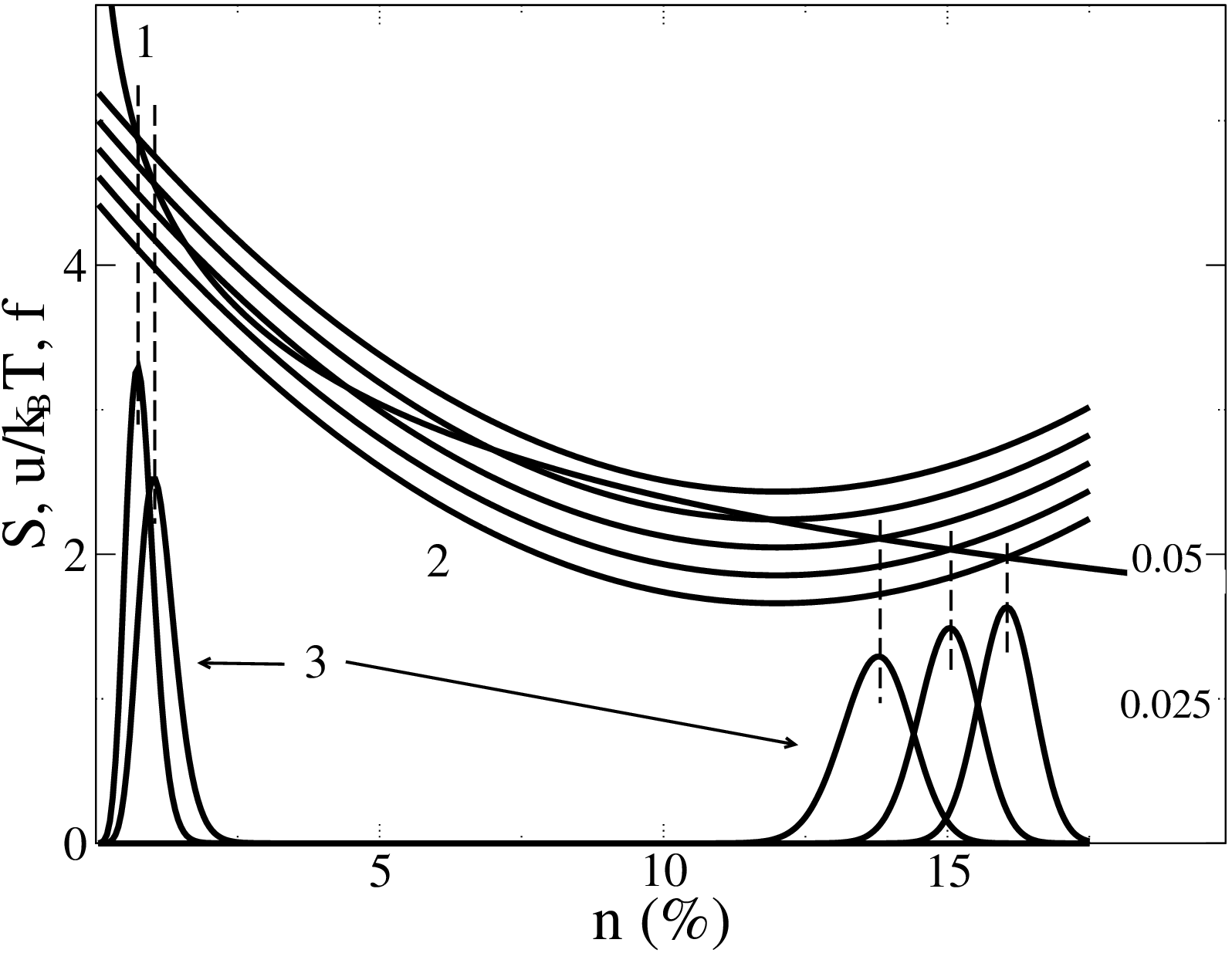}
\caption{\label{f2} Dependence of fundamental curve $S$ (1), energy of vacancy $u$ (2) and $PDF$ (3) on the number of vacancies $n$ at different values of parameter $u_{0}: 0.135; 0.13; 0.125; 0.12; 0.125 (eV)$. The rest parameters are $u_{1} = 0.6\cdot10^{-3} (eV)$, $u_{2} = 0.125\cdot10^{-5} (eV)$, $T=300K$. The right-hand scale is for PDF.}
\end{figure}

The first two terms are sums $S$ of the a slowly divergent harmonic series. This part of the equation depends only on system size and doesn't depend on material parameters. It is the fundamental part of Eq. \ref{b8} decreasing with the growing parameter $n$ (curve 1 of Fig. \ref{f2}). The last term in Eq. \ref{b8} depends on material parameters through coefficients $u_{k}$. At different relations between these coefficients Eq. \ref{b8} may have several solutions. For example, in cubic approximation, Eq. \ref{b8} may have sole solution (the uppermost curve and the two lower curves of series 2 in Fig. \ref{f2}) or two stable solutions (the two intermediate curves) with dependence on parameter $u_{0}$.  To calculate the fundamental curve $S$ the value $N=2000$ was chosen.  As seen, the positions of routes of Eq. \ref{b8} coincide with maxima of PDF calculated directly by Eqs. (\ref{b1})-(\ref{b4}).

The second upper curve of series 2 in Fig. \ref{f2} corresponds to coefficient $u_{0}$ of numerical value $\sim0.15 eV$. For many real materials the energy of vacancy formation is higher, for example, in the case of cooper it is about $1 eV$, where Eq. \ref{b8} has the single solution. Nevertheless, this coefficient can be reduced at the cost of tensile stresses \cite{bsls69,gl97}, size effects \cite{qw03} or a temperature rise \cite{rnzs92,zns93,kl05}, and, as a result, a region with two solutions of Eq. \ref{b8} can be reached. In this case, a structural phase transition between two stable states, one of which is in the region of low defectiveness, and the other in the region of high defectiveness, is possible. The kinetics of such structural phase transition will be considered below.

From the table value of partial sums one can find \cite{gr07}
  \begin{equation}\label{b9}
\sum_{k=1}^{n}\frac{1}{k}=C+\ln n +\frac{1}{2n},
  \end{equation}
where $C$ is a constant. By substituting (\ref{b9}) into (\ref{b8}), for $N>>n>>1$ we obtain a relation between the number of defects $n$ and average energy of defect $u$ in the equilibrium state
  \begin{equation}\label{b10}
n=N\exp(-\frac{u}{kT}).
  \end{equation}

As evident from the last formula, the energy of defect is not strictly constant, but it depends on the total number of defects. Relation (\ref{b10}) is an equation of state for the equilibrium case, relation (\ref{b7a}) is the equation of state too, but for a more general nonequilibrium case including the equilibrium state as a partial case. Eqs. (\ref{b7a}) and (\ref{b10}) need be considered together as a set of equations for deducing both the energy of defect $u_{e}$ and the density of defects $n_{e}$ in the equilibrium state. 

Thus, the equation of state (\ref{b10}) is obtained from condition of the most probable state as the maximum of probability distribution function (\ref{b1}). Same result can be obtained from the principle of free energy minimum. Really,   
the above-described procedure can be schematically displayed as \cite{s89}
  \begin{equation}\label{b19}
U(n)-kT\ln W\equiv U-TS_{c}=F_{c}\rightarrow min.
  \end{equation}

It need be mentioned that product $TS_{c}$ introduced into the definition of the canonical free energy $F_{c}$ is a bound energy lost by the system for doing work. On the other hand, the total energy of defects in the main part is physically the energy lost for the production of work too. Only a little part of it remains for the work production. It follows that
  \begin{equation}\label{b20}
TS_{c}\approx un.
  \end{equation}

And now we can introduce a new specific kind of the free energy by subtracting the bound energy in the form of product $un$ from the internal energy (\ref{b4}). In quadratic approximation (Fig. \ref{f1})
  \begin{equation}\label{b21}
\tilde{F_{c}}=U-un=U_{0}+\dfrac{1}{2u_{1}}(u_{0}-u)^{2}.
  \end{equation}
Here we use the equation of state (\ref{b7a}) to eliminate the density of defects. It is easily found that
  \begin{equation}\label{b22}
n=-\frac{\partial \tilde{F_{c}}}{\partial u}.
  \end{equation}
The both relations (\ref{b7a}) and (\ref{b22}) are a connected couple of equations between the internal energy $U$ and the modified configurational free energy $\tilde{F_{c}}$ on one hand, and between the density of defects $n$ and the energy of defect $u$ on the other hand. One can see that the density of defects is the eigen-argument for the internal energy, and the energy of defect is the eigen-argument for the modified configurational free energy. Hence, the exact free energy $F_{c}$ is expressed according to (\ref{b2}) and (\ref{b19}) through variable $n$, which is not its eigen-argument.

Because the energy needed for the formation of a new defect is smaller, in the presence of others than, in defect-free crystal, the quadratic term in (\ref{b4}) has negative sign. Note that expression (\ref{b4}) is true for both equilibrium and non-equilibrium states. In this approximation the internal energy is a convex function of the defect number having the maximum at point $n=n_{max}$, as shown in Fig. \ref{f1} a. In the same approximation the modified configurational free energy is a concave function with the minimum at point $u=u_{0}$.

With relationships (\ref{b7a}) and (\ref{b10}) it is easy to show that the stationary state corresponds neither to the maximum of the internal energy nor to the minimum of the free energy $\tilde{F_{c}}$. The stationary state is at point $n=n_{e}$, where 
  \begin{equation}\label{b23}
u_{e}=\dfrac{\partial U}{\partial n_{e}},  
\quad n_{e}=-\dfrac{\partial \tilde{F_{c}}}{\partial u_{e}}.
  \end{equation}
Here the additional subscript $e$ denotes the equilibrium value of a variable.

If the system has deviated from the stationary state it should tend back to that state at a speed the higher, the larger the deviation \cite{m07a, m09, m10}
  \begin{equation}\label{b24}
\dfrac{\partial n}{\partial t}=\pm\gamma_{n}(\dfrac{\partial U}{\partial n}-u_{e}),  
\quad \dfrac{\partial u}{\partial t}=\mp\gamma_{u}(\dfrac{\partial \tilde{F_{c}}}{\partial u}+n_{e}).
  \end{equation}

The form of kinetic equations (\ref{b24}) is symmetric with respect to the use of internal and configurational free energy. If the equilibrium state is closer to the maximum of internal energy, then in view of stability of the solution the upper sign is chosen (convex function), if it is near the minimum of internal energy, the lower sign is taken (concave function).  The both variants of the kinetic equations are equivalent and their application is a matter of convenience.

In the right side of the well-known Landau-Khalatnikov kinetic equation \cite{lh54}
  \begin{equation}\label{b25}
\dfrac{\partial n}{\partial t}=-\gamma \dfrac{\partial F_{c}}{\partial n}
  \end{equation}
the ``chemical potential'' is in the form
  \begin{equation}\label{b26}
\mu =\dfrac{\partial F_{c}}{\partial n}.
  \end{equation}

From the thermodynamic point of view, a variable of such kind is not chemical potential really, as is specified by the ``argument'' foreign to the free energy. Still, this notion can be used in practical work, as it directly realizes the minimization principle for the canonical free energy.

If we assume that the equilibrium energy of defect $u_{e}$ and the number of defects $n_{e}$ are slowly changing during an external action then their product can be introduced under differentiation sign in (\ref{b24}) and a new kind (shifted) of internal and free (effective) energies can be defined
  \begin{equation}\label{b27}
\bar{U} =U-u_{e}n,  
\quad \bar{F_{c}} =\tilde{F_{c}}+un_{e}.
  \end{equation}

Then equations (\ref{b24}) are simplified a little
  \begin{equation}\label{b28}
\dfrac{\partial n}{\partial t}=\pm\gamma_{n}\dfrac{\partial \bar{U}}{\partial n},  
\quad \dfrac{\partial u}{\partial t}=\mp\gamma_{u}\dfrac{\partial \bar{F_{c}}}{\partial u}.
  \end{equation}

The original potentials $U$ and $\tilde{F_{c}}$ are connected by means of a Legendre-like transformation
  \begin{equation}\label{b29}
\tilde{F_{c}}=U-un.
  \end{equation}

The shifted potentials $\bar{U}$ and $\bar{F_{c}}$ are connected by means of transformation
  \begin{equation}\label{b30}
\bar{F_{c}}=\bar{U}-un+u_{e}n+un_{e},
  \end{equation}
which differs from the Legendre-like transformation by anticommutation bracket $[un]=u_{e}n+un_{e}$.

\begin{figure}
\hspace{0.06 cm}
\includegraphics [width=2.5 in] {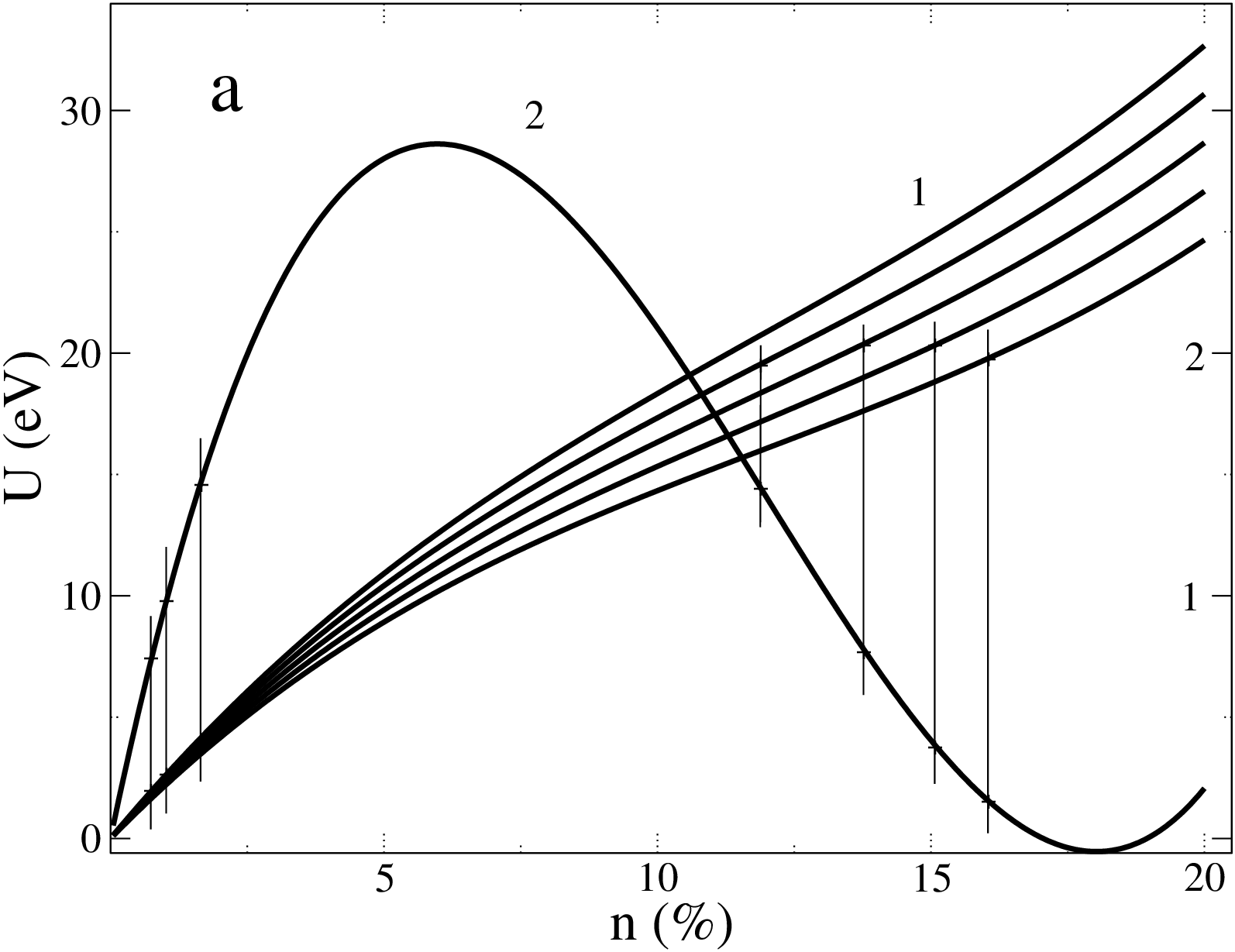}
\vspace{0.09 cm}
\includegraphics [width=2.5 in] {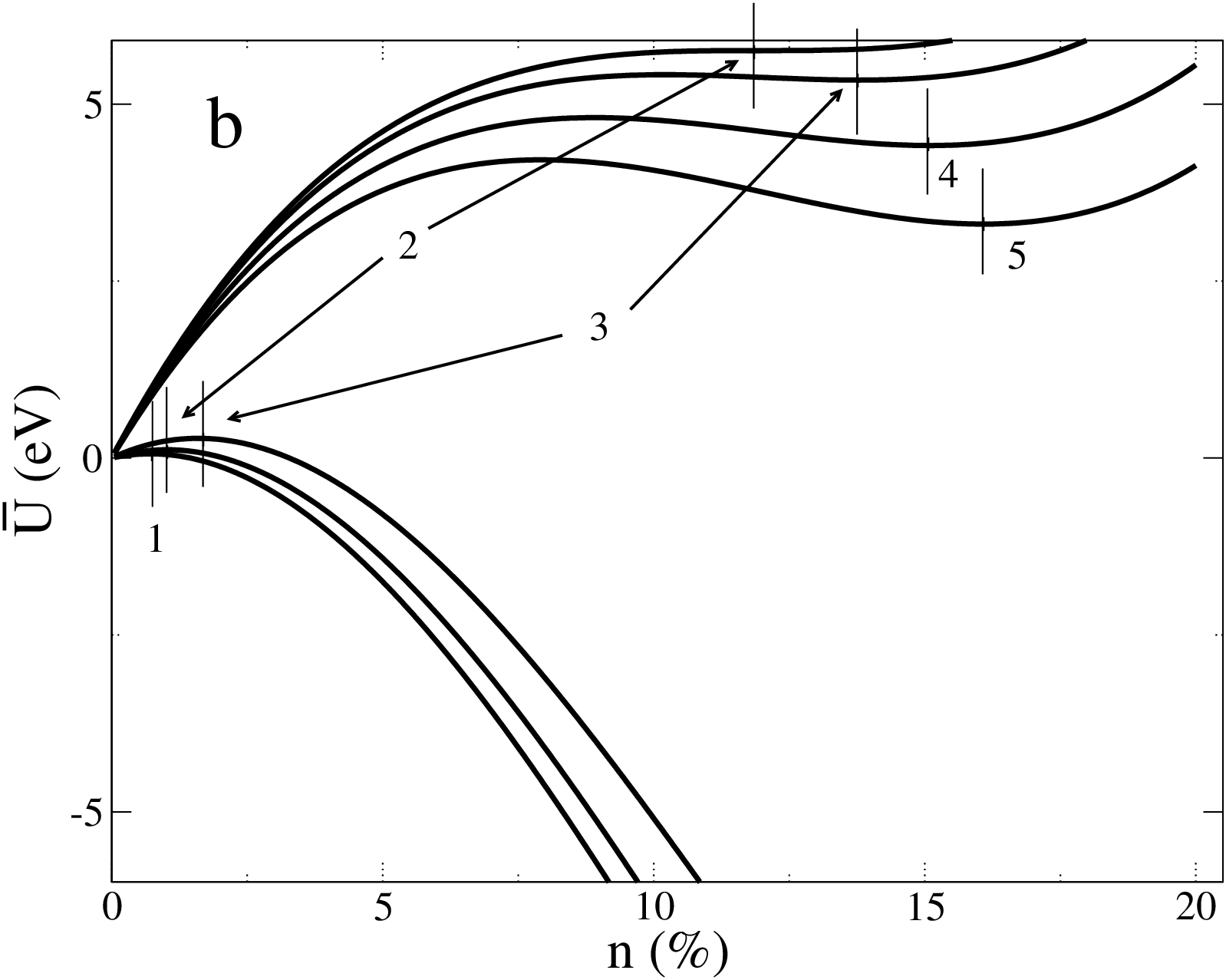}
\caption{\label{f3} Original (a) and effective (b) internal energies as functions of the number of defects: a) $1$ is family of original internal energy; $2$ is linear transform of internal energy with same curvature; b) $1-5$ are positions of the equilibrium states for the same parameters as in Fig. \ref{f2}}
\end{figure}

Graphs of the original and effective internal energies are given in Fig. \ref{f3}. In the considered interval the original internal energy has no extremes, but its curvature changes the sign. In the region of a small number of vacancies it is convex, for a larger number it is concave (curve 2).  That's why the effective internal energy has maximum in the first case and minimum in the second.

The stationary point for the shifted potentials coincides with maximum of $\bar{U}$ (Fig. \ref{f3}) and with minimum of $\bar{F_{c}}$ for the upper signs in (\ref{b28}). Thus, $\bar{U}$ is an effective thermodynamic potential, for which the tendency of the original part of internal energy to minimum is completely compensated by the entropic factor. Twice modified configurational free energy $\bar{F_{c}}$ tends to minimum, but this tendency differs from the case of the canonical configurational free energy $F_{c}$. The effective thermodynamic potential $\bar{F_{c}}$ tends to minimum in the space of eigen-argument $u$, while the canonical free energy $F_{c}$ tends to minimum in the space of non-eigen ``argument'' $n$.


In the paper, a phenomenological approach based on generalization of the Landau technique is considered. For fast processes thermal fluctuations have no time to exert essential influence and it becomes possible to consider the problem in the mean-field approximation. The approach is based not on an abstract order parameter but on physical parameters of structural defects -- their quantity (density) and average energy. The new more general form of kinetic equations, symmetric with respect to using the internal energy $U$ and the modified configurational free energy $\tilde{F_{c}}$, is proposed. In this case, the density of defects and defect energy are related by symmetric differential dependences of type (\ref{b7a}), (\ref{b22}) and (\ref{b23}). As the defect energy in the stationary state is not zero, the extreme principle of equality to zero of the derivative of free energy with respect to ``order parameter'' in the framework of nonequilibrium evolution thermodynamics breaks down. This principle need be substituted with the principle of the tendency to a stationary state. Stationary-state characteristics can not be determined in the framework of phenomenological approach, statistical and microscopic approaches are required.

The present form of kinetic equations can be generalized to all types of regular or randomly distributed defects.

\begin{acknowledgments}
The work was supported by the budget topic № 0106U006931 of NAS of Ukraine and partially by the Ukrainian state fund of fundamental researches (grant F28.7/060).
\end{acknowledgments}

\end{document}